\documentclass[epjCONF]{svjour}
\usepackage{graphics}
\usepackage[varg]{txfonts} 
\usepackage[latin1]{inputenc}
\usepackage[square,sort]{natbib}
\session-title{Conference Title, to be filled}
\begin{document}
\title{GAUFRE: a tool for an automated determination of atmospheric parameters from spectroscopy}
\author{M. Valentini\inst{1}\fnmsep\thanks{\email{valentini@astro.ulg.ac.be}} \and T. Morel\inst{1} \and A. Miglio\inst{2} \and L. Fossati\inst{3} \and U. Munari\inst{4} }
\institute{Institute d'Astrophysique et de G\'{e}ophysique, Universit\'{e} de Li\'{e}ge, B-4000 Li\'ege, Belgium \and School of Physics and Astronomy, University of Birmingham, United Kingdom \and Argelander-Institut f\"ur Astronomie der Universit\"at Bonn, Auf dem H\"ugel 71, 53121, Bonn, Germany \and  INAF-OAPd, Osservatorio Astronomico di Padova, Padova, Italy}

\abstract{
We present an automated tool for measuring atmospheric parameters (T$_{\rm{eff}}$, log~$g$), [Fe/H]) for F-G-K dwarf and giant stars.
The tool, called GAUFRE, is written in C++ and composed of several routines: GAUFRE-RV measures radial velocity from spectra via cross-correlation against a synthetic template, GAUFRE-EW measures atmospheric parameters through the classic line-by-line technique and GAUFRE-CHI2 performs a $\chi^2$ fitting to a library of synthetic spectra. 
A set of  F-G-K stars extensively studied in the literature were used as a benchmark for the program: their high signal-to-noise and high resolution spectra were analyzed by using GAUFRE and results were compared with those present in literature. 
The tool is also implemented in order to perform the spectral analysis after fixing  the surface gravity (log~$g$) to the accurate value provided by asteroseismology. A set of CoRoT stars, belonging to LRc01 and LRa01 fields was used for first testing the performances and the behavior of the program when using the seismic log~$g$.  
} 
\maketitle
\section{Introduction}
Spectroscopy is one of the most powerful tool that astronomy possesses in order to derive atmospheric parameters and abundances of stars. From the stellar spectrum it is possible to measure the effective temperature (T$_{\rm{eff}}$), the surface gravity (log~$g$) and the abundance of iron ([Fe/H]) and of several other elements. \\
The classic method consists in measuring the equivalent widths (EW) of a species in two different ionization states, usually FeI and FeII. By imposing excitation and ionization equilibrium through  stellar atmosphere models, it is possible to derive T$_{\rm{eff}}$, log~$g$ and to infer elemental abundances from the curves of growth. This method is precise and it is widely adopted \cite{Mor2}. In spite of the precision, the line-by-line classical analysis is time consuming: usually EWs are measured by hand using the IRAF ``splot" routine (where the choice of the continuum and the position of the line is completely manual) and the procedure for finding the best T$_{\rm{eff}}$ and log~$g$ requires a large number of iterations.\\
The growing amount of stellar data due to large surveys (i.e. RAVE, SEGUE, LAMOST, HERMES and Gaia ESO) and the availability of dedicated telescopes and multi-object spectrograph, requires the development of automated pipelines, methods and programs that fasten the analysis process. For example, DAOSPEC \cite{Stet} and ARES \cite{Sou} are useful and free codes that performs an automated EW measurement; the MOOG code \cite{Sne} is a valid collection of routines useful for determining atmospheric parameters and abundances. In particular, MOOG \textit{abfind} and \textit{synth} tasks are widely used in literature \cite{Ran} \cite{Mor2}.  \\
Another method for the estimation of atmospheric parameters and elemental abundances is to compute a set of synthetic spectra and to  find the best match between the synthetic and the observed spectrum. This method is adopted by surveys as RAVE \cite{Stei} and HERMES \cite{Free} as well as in small surveys as ARCS \cite{Val} \cite{Sag}.\\
In this paper we present a new automatic code, GAUFRE, that can perform both type of analysis. In section~\ref{sec:1} we give a short description of the idea behind the program and the two types of analysis that it performs. In subsection~\ref{sb:24} we present the first results of tests using spectra of objects well known in literature. In section~\ref{sec:3} we present an additional tool of GAUFRE that consists in using the asteroseismic gravity as a fixed value for log$(g)$ in order to refine the measurement of T$_{\rm eff}$, microturbulence velocity ($\xi_{mic}$) and elemental abundances. Section \ref{sec:4} discusses the future perspectives for the code.\\
\begin{table}
\centering
\caption{Comparison between the radial velocity we measured with GAUFRE-RV (adopting the library of synthetic spectra of Fossati) and IAU values. All radial velocities are in km s$^{-1}$}
\label{tab:1}       
\begin{tabular}{llllllll}
\hline\noalign{\smallskip}
Star & Instrument & &\multicolumn{2}{c}{IAU} & &\multicolumn{2}{c}{GAUFRE-RV}  \\
         &                        & &v$_{rad}$& $\sigma_{v_{rad}}$& & v$_{rad}$ & $\sigma_{v_{rad}}$ \\
\noalign{\smallskip}\hline\noalign{\smallskip}
HD 003712 & Asiago-Echelle&   & $-$3.9  & 0.1 & & $-$4.1  & 0.3    \\ 
HD 012929 & Asiago-Echelle & & $-$14.3 & 0.2 & & $-$13.8 & 0.4    \\      
                      & ESO-FEROS     & &                &        &   & $-$14.6 & 0.2 \\
HD 062509 & Asiago-Echelle&  & $+$3.3  & 0.1 & & $+$3.3  & 0.6     \\
                       & ESO-UVES       & &               &        & & $+$3.4  & 0.3     \\ 
HD 065934  & Asiago-Echelle& & $+$35.0 & 0.3 & & $+$34.6 & 0.5    \\ 
HD 090861 & Asiago-Echelle&  & $+$36.3 & 0.4 & & $+$36.7 & 0.4    \\ 
HD 212943  & Asiago-Echelle& & $+$54.3 & 0.3 & & $+$54.2 & 0.4    \\
                      &ESO-UVES        &  &                &        & &  $+$54.1 & 0.2 \\
HD 213014  &  Asiago-Echelle& & $-$39.7 & 0.0 & & $-$40.0 & 0.5    \\ 
\noalign{\smallskip}\hline
\end{tabular}
\end{table}

\section{The GAUFRE code}
\label{sec:1}
GAUFRE is a collection of several C++ routines. It is written in order to measure fast and precisely radial velocity (V$_{\rm rad}$) and atmospheric parameters (T$_{\rm{eff}}$, log~$g$, [Fe/H])  of a star starting from a one-dimensional normalized spectrum. The radial velocity is measured with a cross-correlation technique (routine GAUFRE-RV), atmospheric parameters can be measured adopting a $\chi^2$ fitting over a library of synthetic spectra (GAUFRE-CHI2 routine) or with the classic FeI-FeII lines technique (GAUFRE-EW routine). \\
GAUFRE is an updated and extended version of the code written for the Asiago Red Clump Spectroscopic survey (ARCS) \cite{Val}. GAUFRE was created because of the need of introducing new libraries of synthetic spectra, adapting the code for different resolutions and implementing the classic line-by-line technique. The code was written in C++ and does not need any particular library, in order to avoid any software license problems and to be executable in different platforms. The only additional program needed for running GAUFRE is MOOG, which can be easily installed and downloaded from its homepage (http://www.as.utexas.edu/$\sim$chris/moog.html). The user is also supposed to download the libraries of synthetic spectra and model atmospheres required by GAUFRE. 
So far GAUFRE is tested for F-G-K stars (both giants and dwarfs) and works in the 3500 - 9000 ~\AA~ spectral range. The program has currently been tested for spectral resolutions of the Asiago Echelle Spectrograph (R=20,000),  ESO-FLAMES-GIRAFFE HR10, HR15N, HR21, HR14 setups (R=16,000 - 20,000), ESO-FLAMES-UVES U580 setup (R=52,000), ESO-FEROS (R=48,000) and AAOmega 2dF 580V and 385R (R=1,300). Details about the adopted libraries can be found on subsections \ref{sb:23} and \ref{sb:24}.

\subsection{Radial Velocities: GAUFRE-RV}
\label{sb:21}
The radial velocity subroutine (GAUFRE-RV) measures the radial velocity V$_{\rm{rad}}$ by cross-correlating the observed spectrum with a synthetic spectral library \cite{Ton}. The procedure is the same as described in \cite{Val}. The procedure starts from the continuum normalized spectrum in a 2-column ASCII format. First, the synthetic normalized spectra of the selected library are renormalized, following the same parameters as adopted for the normalization of the observed spectrum (same function, same order and same high and low rejection values). To lower the impact of the different noise level in the observed and synthetic spectra, they are scaled to match their geometric mean. This procedure is needed in order to improve the accuracy of the cross-correlation and of the $\chi^2$ fitting and it is performed by the GAUFRE-LIB routine (see also section 2.2).\\
The result of the GAUFRE-RV subroutine is a file containing the value of  V$_{\rm{rad}}$ and an ascii file containing the continuum normalized spectrum corrected for the radial velocity.\\
To validate the GAUFRE-RV routine we measured the radial velocity of a set of Red Giants stars that are IAU standard radial velocity stars. These stars were observed with the Asiago Echelle Spectrograph (INAF-OAPd); we then downloaded, when available, the spectrum from the ESO-Archive. For this test, we adopted the synthetic spectra library provided by L. Fossati (cf. section \ref{sb:22}). Results are summarized in Table~\ref{tab:1}.  The mean difference between our values and those present in the literature is $\Delta RV_{\odot}=0.10$ km s$^{-1}$, with a rms of 0.3 km s$^{-1}$. \\
 
\subsection{$\chi2$ on synthetic spectra libraries: GAUFRE-CHI2}
\label{sb:22}
Atmospheric parameters (T$_{\rm{eff}}$, log~$g$, [Fe/H], [$\alpha$/Fe], V$_{\rm rot}\sin\,i$) are obtained by GAUFRE-CHI2 via  $\chi^2$ fitting of the continuum normalized spectrum against a synthetic spectral library. \\
The choice of the spectral library is up to the user, available libraries are: a library based on Kurucz model atmosperes \cite{Mun}, the library provided by L. Fossati  (built adopting the spectral synthesis code Synth3 described in \cite{Koch}) and the AMBRE library \cite{deL}. Characteristics of different libraries are summarized in Table \ref{tab:2}. Libraries are provided at different resolutions and they cover a wide spectral range (usually 3,500 - 10,000~\AA). Before starting $\chi^2$ analysis, the desired library must be cut, normalized and degraded at the same wavelength interval, normalization function and resolution of the real spectra. For this purpose a routine GAUFRE-LIB has been created. The degradation of the spectra at the desired resolution is performed through deconvolution with a Gaussian profile. \\

\begin{table}
\caption{Libraries of synthetic spectra used by GAUFRE-RV and GAUFRE-CHI2.}
\label{tab:2}       
\begin{tabular}{llllllllll}
\hline\noalign{\smallskip}
Library & Model  & Linelist & Solar abundances & Resolution \\
              & atmospheres            &               &                             &                        \\
\noalign{\smallskip}\hline\noalign{\smallskip}
Munari et al. (2005) & Castelli \& Kurucz (2003) & Kurucz & Grevesse \& Sauval (1998) & 20,000   \\
Fossati                       & Castelli \& Kurucz (2003)  & VALD2  &Grevesse \& Sauval (1998) & 1,000,000 \\
De Laverny  et al. (2012)   & MARCS                               &              & Grevesse, Asplund \& Sauval (2007)  & 150,000 \\
\noalign{\smallskip}\hline
\end{tabular}
\end{table}

\subsection{EW and MOOG: GAUFRE-EW}
\label{sb:23}
\begin{figure}
\begin{minipage}[b]{0.45\linewidth}
\centering
\resizebox{1.0\columnwidth}{!}{\includegraphics{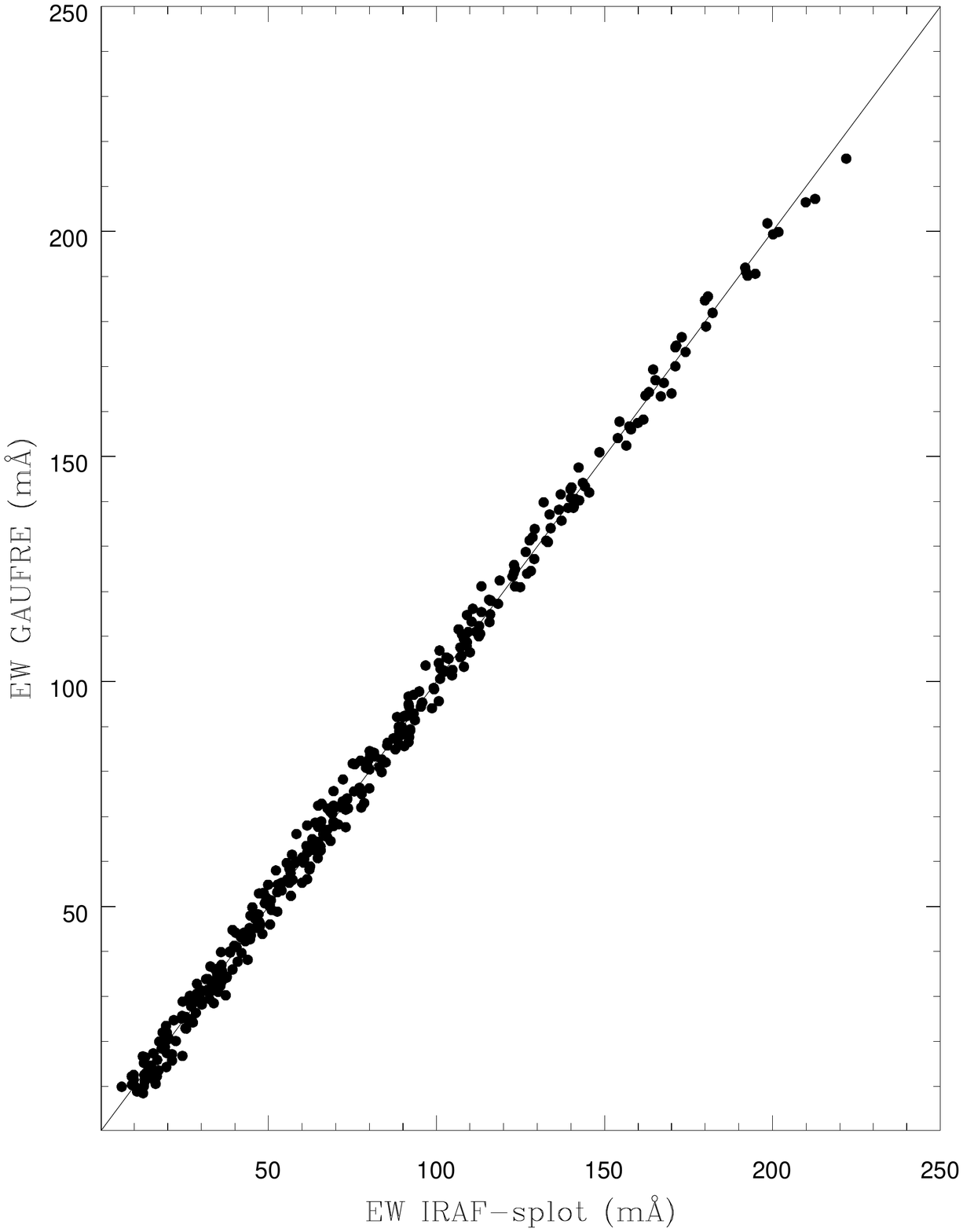} }
\end{minipage}
\begin{minipage}[b]{0.45\linewidth}
\centering
\resizebox{1.1\columnwidth}{!}{\includegraphics{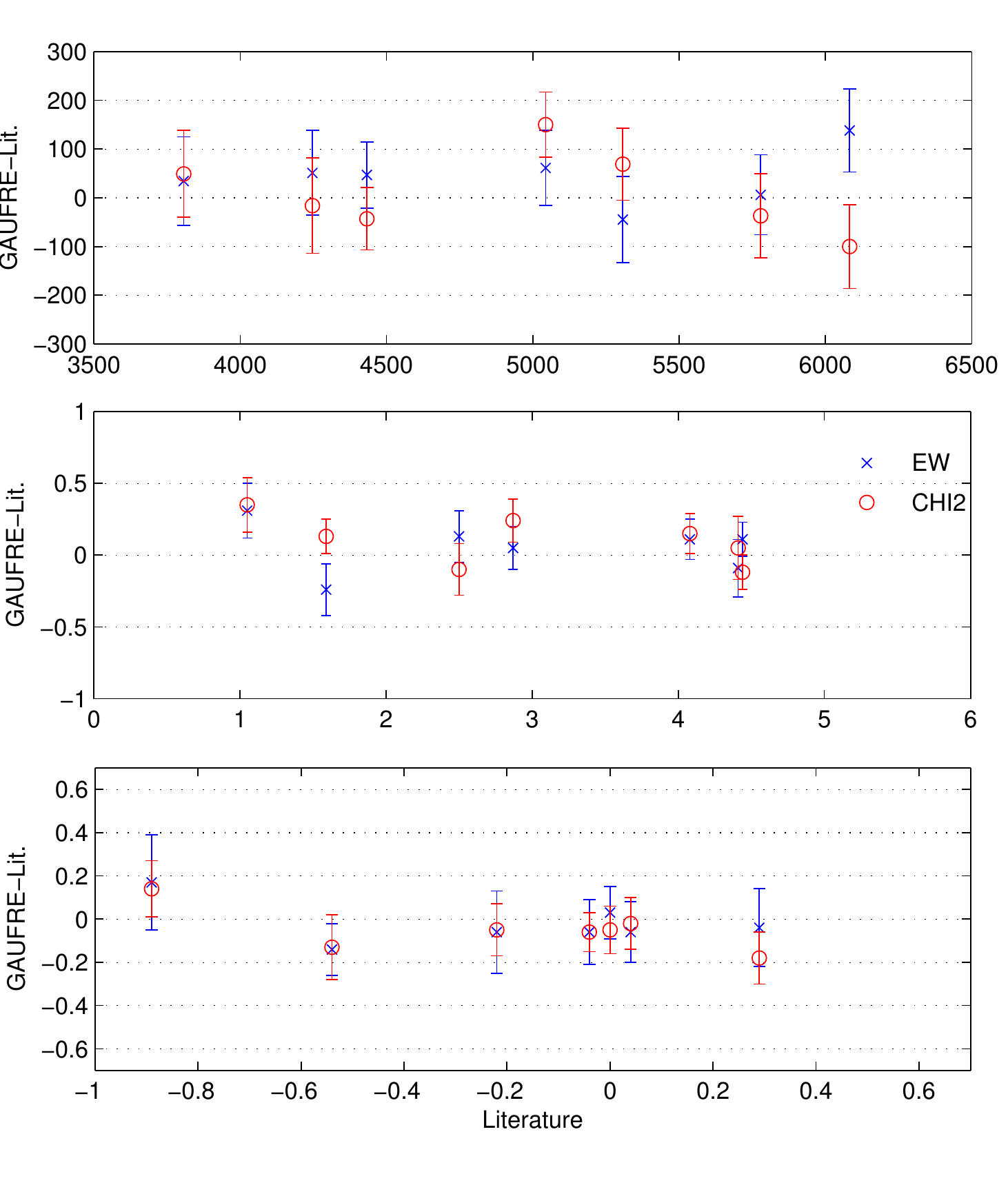} }
\end{minipage}
\caption{Left panel: Comparison between the EW measured by GAUFRE-EW (y-axis) and the EW measured with the IRAF-splot routine, fitting the line to a Gaussian profile. Right panel: Differences between the values of 
T$_{\rm eff}$ (top panel), $\log\,g$ (middle panel) and [Fe/H] (bottom panel) measured by GAUFE-EW and GAUFRE-CHI2 and those present in literature.}
\label{fig:1}       
\end{figure}

GAUFRE-EW automatically performs the classical line-by-line analysis for deriving atmospheric parameters and abundances. The procedure starts from a continuum normalized spectrum (ASCII 2 columns format), a file containing a list of lines to measure and their parameters (as requested by MOOG) and a file containing the parameters of the spectrum like the wavelength coverage, resolution and, if any, guessed values for T$_{\rm{eff}}$ and log~$g$.\\
The EW of every line present in the input file is measured (except when it is not detectable). The program selects an area of 3-4 \AA~ around the wavelength of the line (this parameter is selected by the user). The spectrum is then fitted with a polynomial function in order to determine the continuum and point with the lowest intensity. In order to test the automated EW measurement of the line we compared the EW values obtained by measuring features by hand (using IRAF-splot) and the correspondent EW measured by GAUFRE. The test was performed on the spectrum of Arcturus, taken with the ESO-FLAMES-UVES instrument, with the U580 setup (5770-6825 \AA). The agreement is quite good and rms is of $\sim$3 m\AA ~(see Figure \ref{fig:1}).\\
Atmospheric parameters were computed by using MOOG \textit{abfind} driver (the program uses its non-iinteractive version, MOOGSILENT), the measured EW of FeI and FeII lines and a family of model atmospheres (MARCS \cite{Gus} or Kurucz \cite{Cas}). T$_{\rm{eff}}$ is calculated by assuming the excitation equilibrium and minimizing the trend of the Fe abundance  versus the excitation potential. The surface gravity, log~$g$ is derived by assuming the ionization equilibrium: log~$n$~(FeI)= log~$n$~(FeII). The procedure is iterative and the program will converge to  T$_{\rm{eff}}$ and log~$g$ that satisfy both the ionization and excitation equilibria. The value of the microturbulence $\xi_{\rm{mic}}$ is derived by minimizing the trend of the FeI abundance versus the FeI lines EW.\\

\subsection{Atmospheric parameters validation}
\label{sb:24}
We took spectra of 7 F-G-K stars taken with FLAMES-UVES U580 or FEROS from the ESO-archive (http://archive.eso.org). We selected spectra of targets very well known in literature: $\alpha$~Cen A, $\mu$~Cas A, $\beta$~Vir, Arcturus, $\mu$~Leo, $\xi$~Hya and $\gamma$~Sge. As a reference we used an average value of the most recent entries of the PASTEL catalog \cite{Sou}. We analyzed spectra with both GAUFRE-EW (Kurucz model atmospheres) and GAUFRE-CHI2 (Fossati library of synthetic spectra) and we compared our results with those present in literature. The agreement is quite good as showed in Tab.3.\\
\begin{table}
\caption{Comparison of the atmospheric parameters obtained by GAUFRE-CHI2 (G-CHI2) and GAUFRE-EW (G-EW) with thopse present in literature for a set of 7 stars.}
\label{tab:2}       
\begin{tabular}{llll}
\hline\noalign{\smallskip}
\multicolumn{2}{l}{GAUFRE-CHI2}&\multicolumn{2}{l}{GAUFRE-EW}\\
\noalign{\smallskip}\hline\noalign{\smallskip}
$\langle T_{\rm eff}|_{\rm G-CHI2}-T_{\rm eff}\rangle=16$~K &  $\sigma=84$~K & $\langle T_{\rm eff}|_{\rm G-EW}-T_{\rm eff}\rangle=0$~K & $\sigma=17$~K\\
$\langle \log g|_{\rm G-CHI2}-\log g\rangle=0.13$~dex &  $\sigma=0.17$~dex & $\langle \log g|_{\rm G-EW}-\log g\rangle=0.11$~dex & $\sigma=0.17$~dex\\
$\langle {\rm [Fe/H]}|_{\rm G-CHI2}-{\rm [Fe/H]}\rangle$=-0.05~dex &   $\sigma=0.10$~dex & $\langle {\rm [Fe/H]}|_{\rm G-EW}-{\rm [Fe/H]}\rangle=$-0.06~dex & $\sigma=0.10$~dex\\
\noalign{\smallskip}\hline
\end{tabular}
\end{table}

\section{Asteroseismic constraints}
\label{sec:3}
\begin{figure}
\resizebox{1.0\columnwidth}{!}{\includegraphics{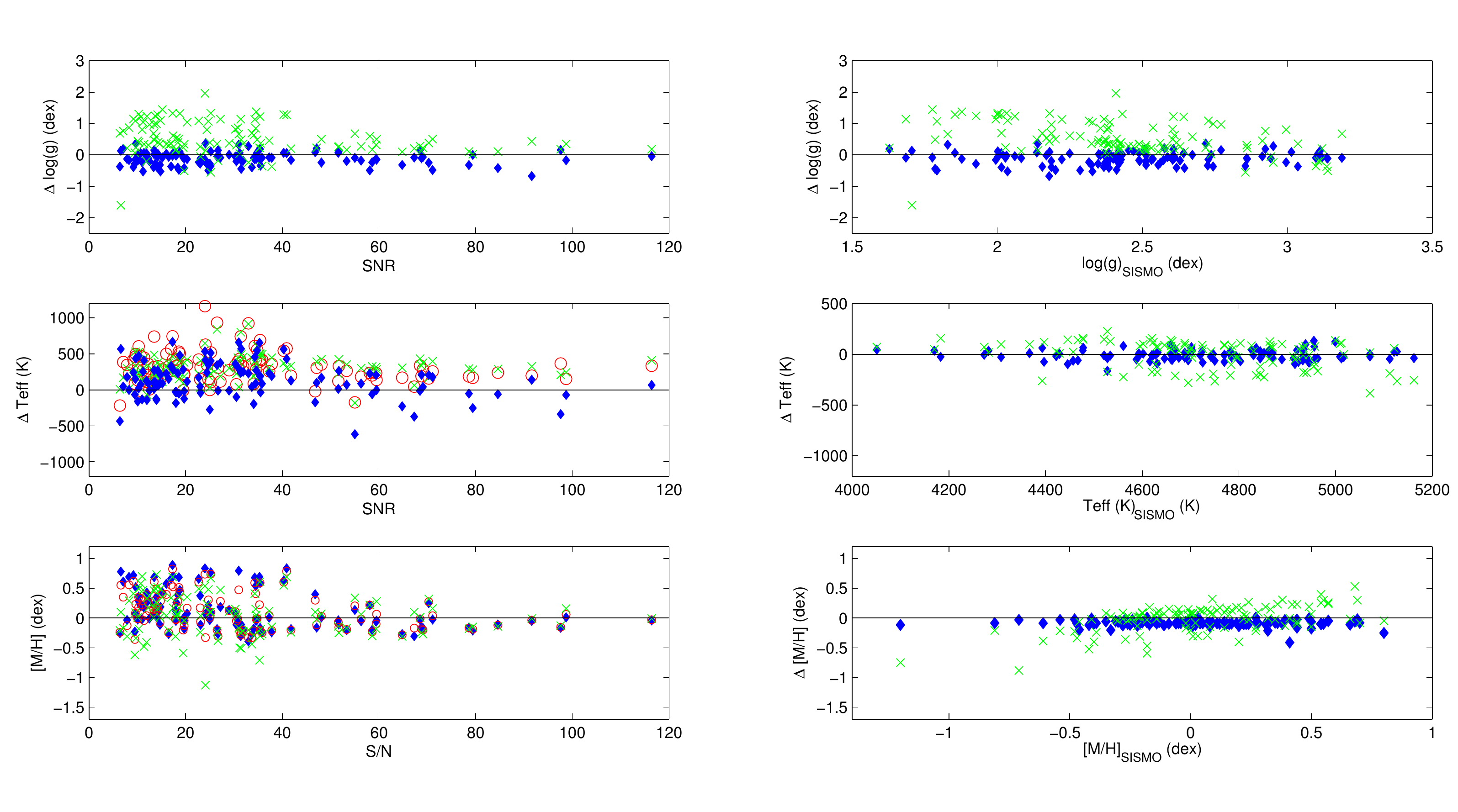}}
\caption{Left panel: differences between the atmospheric parameters measured with different techniques (GAUFRE-CHI2 this paper, MATISSE \cite{Gazz}, photometry  $(J-K)$ and asteroseismology) in function of the SNR. Right panel: differences between the atmospheric parameters measured with different techniques (GAUFRE-CHI2 this paper, MATISSE by \cite{Gazz}, photometry $(J-K)$ and asteroseismology) in function of the T$_{\rm{eff}}$, log $g$ and [M/H] measured by GAUFRE-CHI2 by fixing the log~$g$ to the seismic value.  Green crosses are data from \cite{Gazz}, blue diamonds are data obtained by using GAUFRE-CHI2 and red circles are values measured by GAUFRE-CHI2 by fixing the log~$g$ to the seismic value. It is worth to take into account the poor quality of the spectra: more than 50\% of the spectra possess a SNR $<$ 40.}
\label{fig2}       
\end{figure}

As extensively discussed, for example, in \cite{Mor1}, the frequency of maximum power, $\nu_{\rm max}$ can be used for deriving a precise estimate of the surface gravity:\\
\begin{equation}
\log g = \log g_\odot + \log \left(
{{\nu_{\rm max} \over \nu_{{\rm max}, \, \odot}}}
\right)
+ {1 \over 2}
\log \left({T_{\rm eff}\over {\rm T}_{{\rm eff}, \, \odot}}\right) {\rm .}
\label{eq:1}
\end{equation}
The precision of the gravity derived from Eq. \ref{eq:1} is generally expected to be below 0.05~dex, thanks to the weak sensitivity of the scaling relation to the assumed T$_{\rm{eff}}$ and the high precision usually achieved for the measurement of $\nu_{\rm max}$.\\
The scaling relation has been demonstrated to be reliable \cite{Bru} \cite{Stel} \cite{Mig} \cite{Hub} and it can be used for refining atmospheric parameters and abundances. As a matter of fact one can improve the spectroscopic analysis by fixing the $\log g$ to the seismic value: this largely increases the accuracy of the derived $T_{\rm eff}$, [Fe/H] and hence chemical abundances.\\
A set of spectra of 111 RG stars belonging to the LRc01 and LRa01 fields of CoRoT have been used as benchmark for testing the log~$g$ values derived by GAUFRE. Spectra are taken with the ESO-FLAMES GIRAFFE 9 setup, centered in the MgI triplet (5278 \AA). Spectra have already been analyzed \cite{Gazz} by using MATISSE tool \cite{Recio}. \\
First, we compared $\log g$ values of GAUFRE-CHI2 (using the synthetic library provided by L. Fossati, see Tab\ref{tab:2}) with those provided by asteroseismology (see Fig\ref{fig2}, top panels), then we compared the values of T$_{\rm{eff}}$ and [Fe/H] derived by fixing $\log g$ to the seismic value (see Fig\ref{fig2}, middle and bottom panels). The set of spectra taken from the work of Gazzano has very poor SNR (see Fig\ref{fig2}), of $\sim$30 on average. In addiction the spectral range covered by spectra is affected by the presence of MgH molecular bands. This complicates even more the continuum normalization, leading to several systematics in the atmospheric parameters determination. 

As shown in Fig.~\ref{fig2}, the seismic gravity can be used to enhance the accuracy of the atmospheric parameters. In the case of Gazzano et al. dataset \cite{Gazz} with a poor SNR, the seismic gravity helped in providing a more precise determination of T$_{\rm{eff}}$ and [M/H] reducing average errors from 120 K and 0.20 dex to 75 K and 0.11 dex respectively.

\section{Conclusions}
\label{sec:4}
We present the GAUFRE program,  a versatile tool developed for measuring  radial velocities and atmospheric parameters from optical spectra.\\
The program is composed by different routines: GAUFRE-RV, GAUFRE-CHI2, GAUFRE-EW.\\
We performed some preliminary tests in order to check the performances of the tool. These tests show that the program is reliable and that it can be used to process spectra of F-G-K dwarfs and giants with a SNR above 40. At the moment it is adopted by the Li\`ege node within the Gaia-ESO Survey (PI: G. Gilmore and S. Randich) and further applications are planned.\\
We also showed an interesting and useful extension of GAUFRE that uses, when avaiable, the seismic log~$g$ as a fixed value for the surface gravity: the precise and also likely more accurate values given by asteroseismology allow us to greatly refine the values of T$_{\rm{eff}}$ and [Fe/H].\\
The GAUFRE tool is continously in development: we plan to implement new libraries of synthetic spectra and to extend the analysis to the infrared region. New and detailed tests are planned as well, in order to better investigate the performances of GAUFRE at different SNR.\\
In the next future we plan to make the GAUFRE code avaiable through the web and to create an user-friendly graphical interface.\\
\\

\footnotesize{MV aknowledges financial support from Belspo for contract PRODEX COROT. TM acknowledges financial support from Belspo for contract PRODEX GAIA-DPAC. Paper based on observations collected at Asiago Observatory and from data recovered from the ESO Archive.}
\bibliographystyle{epj}
\bibliography{ValentiniM}

\end{document}